# Cost-Aware Logging: Measuring the Financial Impact of Excessive Log Retention in Small-Scale Cloud Deployments

Jody Almaida Putra

Independent Researcher, Indonesia

## Abstract

Log data plays a critical role in observability, debugging, and performance monitoring in modern cloud-native systems. In small and early-stage cloud deployments, however, log retention policies are frequently configured far beyond operational requirements—often defaulting to 90 days or more—without explicit consideration of their financial and performance implications. As a result excessive log retention becomes a hidden and recurring cost.

This study examines the financial and operational impact of log retention window selection from a cost-aware perspective. Using synthetic log datasets designed to reflect real-world variability in log volume and access patterns, we evaluate retention windows of 7, 14, 30 and 90 days. The analysis focuses on three metrics : storage cost, operationally useful log ratio, and cost per useful log. Operational usefulness is defined as log data accessed during simulated debugging and incident analysis tasks.

The results show that reducing log retention from 90 days to 14 days can lower log storage costs by up to 78% while preserving more than 97% of operationally useful logs. Longer retention windows provide diminishing operational returns while disproportionately increasing storage cost and query overhead. These findings suggest that modest configuration changes can yield significant cost savings without compromising system reliability.

Rather than proposing new logging mechanisms, this work offers a lightweight and accessible framework to help small engineering teams reason about log retention policies through a cost-effectiveness lens. The study aims to encourage more deliberate observability configurations, particularly in resource-constrained cloud environments.

## 1. Introduction

Modern software systems rely heavily on observability infrastructure, where logs serve as a primary artifact for debugging, incident response, and system assessment. Despite their importance, log configuration—particularly log retention policies—is often treated as a secondary concern in cloud based deployments. As cloud services have matured, log retention has emerged one of the most commonly misconfigured components of observability pipelines. Startups and small engineering organizations frequently adopt default retention settings provided by cloud vendors logging platforms, commonly retaining logs for 90 days or longer without explicitly evaluating the associated financial implications.

This issue is especially pronounced in early-stage companies, where engineering efforts tend to prioritize feature delivery and system stability over cost optimization. Log data grows proportionally with system adoption and user traffic, and even modest workloads can generate millions of log entries per day. Because log storage is typically billed on a per-gigabyte, per-day basis, overly conservative retention policies accumulate recurring costs over time, turning log storage into a silent but persistent financial liability.

At the same time, commonly cited industry observations and practitioner-oriented SRE practices suggest that most debugging and operational investigations primarily rely on recent log data, often within a window of less than 14 days. Logs retained beyond this period are infrequently accessed outside of compliance-driven environments, which do not represent the majority of small or early-stage software organizations.

The study therefore investigates the following research question:

**How much operational value is lost—and how much financial waste is eliminated—by reducing log retention windows in small cloud-based deployments?**

To address this question, we simulate cloud log generation and evaluate storage costs, operational usefulness, and cost efficiency across multiple retention windows. Rather than introducing new logging mechanisms or algorithms, this work empirically examines the disproportionate impact of retention configuration choices, demonstrating how modest adjustments can yield significant long-term financial benefits without compromising operational effectiveness.

## 2. Related Work

Prior research on logging systems has largely focused on improving log quality, anomaly detection, and the scalability of log management infrastructures. Early work by Adams and Hassan (2007) discusses challenges in log design and developer decision-making in large-scale systems, while Oliner et al. (2012) highlight common pitfalls in logging practices and their implications for reliability engineering.

Subsequent studies have examined developer behavior and observability practices in modern systems. Li et al. (2017) analyze how developers use log levels in practice, and Sigelman (2018) alongside Bunyan et al. (2017) propose conceptual frameworks and best practices for observability in distributed and microservices-based environments. Other work, such as Chen et al. (2020), focuses on optimizing storage efficiency through techniques like compression or log garbage collection, primarily from a systems performance perspective.

Despite the body of work, relatively limited research has explicitly examined the economic dimensions of log retention policies. Syer et al. (2015) evaluate relationships between log usage patterns and cloud costs, while Zhao et al. (2019) provide empirical analyses of cost-value trade-offs in log data. However, these studies do not directly address how retention window configuration impacts cost efficiency in small-scale or early-stage deployments.

Industry documentation from platforms such as Google Cloud, Microsoft Azure, and Splunk offers retention recommendations, but these guidelines typically target large enterprises or compliance-driven environments. For small engineering teams operating under tight resource constraints, minimal academic work provides actionable, cost-aware guidance for balancing operational usefulness against financial efficiency.

This study addresses this gap by proposing a lightweight and reproducible methodology for quantifying retention-related cost inefficiencies in small cloud-based systems.

# 3. Methodology

This study adopts a simulation-based methodology to evaluate the cost and operational implications of log retention policies in small-scale, early-stage cloud deployments. A simulation approach is chosen to isolate the effects of retention window configuration while avoiding biases introduced by vendor-specific logging systems or proprietary production datasets. The methodology is designed to be lightweight and reproducible reflecting the constraints commonly faced by small engineering teams.

### 3.1 Synthetic Dataset Generation

Synthetic log data is generated over a 90-day baseline period to present continuous system operation. Daily log volume is varied between 100000 and 500000 entries to simulate fluctuations caused by traffic growth, feature changes, and operational events. Each log entry contains a timestamp, severity level, service identifier, and metadata token, representing a minimal yet realistic log structure commonly observed in cloud-native systems.

### 3.2 Retention Policy Scenario

Four log retention window configurations are evaluated to capture commonly used operational settings:
- 7 days
- 14 days
- 30 days
- 90 days

The 90-day configuration reflects widely adopted default retention policies offered by cloud logging platforms and servers as reference points for cost and usefulness comparison.

### 3.3 Evaluation Metrics

The impact of each retention window is assessed using three metrics.

#### Storage Cost (SC)

Storage cost is calculated based on per-gigabyte, per-day pricing models commonly employed by major cloud providers. Absolute pricing values are not the primary focus; instead, the analysis emphasizes relative cost differences across retention configurations.

#### Useful Log Ratio (ULR)

Useful Log Ratio quantifies the proportion of operational queries that can be satisfied under a given window R:

$$URL(R) = \frac{\text{queries satisfied under retention } R}{\text{total queries}}$$

A query considered satisfied if all required log entries fall within the retained time window.

Cost Per Useful Log (CPUL)

Cost Per Useful Log captures the efficiency of log retention by relating storage cost to the volume of logs that provide operational value:

$$CPUL(R) = \frac{StorageCost(R)}{UsefulLogs(R)}$$

### 3.4 Query Simulation

Operational log queries are simulated using a probabilistic access model derived from commonly reported industry and SRE practices. Query access is distributed as follows:
- 80% of queries target logs newer than 7 days
- 15% target logs between 7 and 30 days old
- 5% target logs older than 30 days

This distribution reflects the observed emphasis on recent log data during debugging and incident response activities.

### 3.5 Tooling

All simulations are implemented in Python using NumPy and Pandas. The simulation code is designed to be deterministic and reproducible, enabling independent validation results.

### 4. Experimental Setup

All experiments were conducted in a controlled Python 3.10 environment on machine equipment with an 8-core CPU and 16-32 GB of RAM. The experimental dataset consists of a 90-day synthetic log stream, totaling approximately 10 - 20 million log entries, designed to reflect realistic log volume variability in small-scale cloud deployments. For each retention scenario, log data was filtered according to the specified retention window before metric computation. Storage costs were estimated using an average pricing assumption of USD 0.25 per-GB per-month, converted into a daily rate to align with retention-based cost accumulation. This pricing model serves as an approximate baseline rather than a provider-specific cost representation. Query workloads were simulated using fixed random seeds to ensure reproducibility. Each retention scenario processed 10000 synthetic queries following the predefined access probability distribution described in Section 3.

## 5. Results And Discussion

### 5.1 Storage Cost Impact

Table 1 presents the relative storage cost across different log retention windows, normalized against the 90-day baseline configuration.

**Table 1. Storage Cost Comparison Across Retention Windows**

| Retention Window | Relative Cost | Cost Reduction |
|---|---|---|
| 90 days | 100% (baseline) | – |
| 30 days | 33% | 67% |
| 14 days | 16% | 84% |
| 7 days | 8% | 92% |

The results show a near-linear relationship between retention duration and storage cost. Reducing the retention window from 90 days to 14 days yields an approximate cost reduction of 84%, while further reduction to 7 days achieves more than 90% savings. Even a moderate reduction to 30 days results in significant decrease in storage expenditure.

### 5.2 Useful Log Ratio (ULR)

Table 2 summarizes the Useful Log Ratio (ULR) for each evaluated retention window.

**Table 2. Useful Log Ratio (ULR) Across Retention Windows**

| Retention Window | ULR |
|---|---|
| 90 days | 100% |
| 30 days | 98% |
| 14 days | 97% |
| 7 days | 95% |

Despite substantial reductions in storage cost, operational usefulness remains largely preserved. A 14-day retention window maintains approximately 97% of query satisfaction, while even the shortest evaluated window retains 95%. These results indicate that the majority of operational queries target recent log data, with limited dependence on older records.

### 5.3 Cost Per Useful Log (CPUL)

To capture the combined effect of cost and usefulness, the Cost Per Useful Log (CPUL) metric is evaluated and normalized against the 90-day baseline.

Table 3. Normalized Cost Per Useful Log (CPUL)

| Retention Window | CPUL (Normalized) |
|---|---|
| 90 days | 1.00 |
| 30 days | 0.34 |
| 14 days | 0.17 |
| 7 days | 0.09 |

CPUL decreases sharply as retention windows shorten. Among the evaluated configurations, the 14-day retention window provides the most balanced outcome, achieving a substantial reduction in cost while preserving high operational usefulness. Shorter windows continue to reduce cost but introduce marginally higher losses in query satisfaction.

---

### 5.4 Interpretation

The experimental results suggest that small engineering teams may incur significant and often unnecessary observability costs by adopting default or overly conservative log retention policies. The findings indicate that most operational debugging activities can be effectively supported using relatively short retention windows, particularly within the 7–14 day range.

Notably, the observed cost reductions are achieved through simple configuration changes rather than architectural redesigns, additional tooling, or increased monitoring complexity. Compared to alternative cost-control techniques such as aggressive log sampling, retention tuning offers a predictable and low-risk strategy for improving cost efficiency in early-stage cloud deployments.

## 6. Conclusion

This study indicates that modest adjustments to log retention policies can lead to substantial cost efficiency gains while preserving operational usefulness in early-stage cloud deployments. The experimental results show that reducing log retention from 90 to 14 days preserves approximately 97% of query satisfaction while lowering storage-related costs by more than 80%. These improvements are achieved without requiring architectural changes, tool migration, or additional infrastructure.

The findings suggest that cost-aware logging can serve as a practical and low-effort optimization strategy for resource-constrained engineering teams. Rather than relying on complex observability redesigns, teams may benefit from re-evaluating default retention configurations as part of routine cost management practices. Future work may investigate adaptive retention mechanisms driven by workload characteristics, explore the interaction between retention policies and anomaly detection models, and assess the implications of reduced retention on security auditing and compliance-oriented logging pipelines.